\documentclass{article}
\usepackage{graphicx}
\usepackage{amsmath}
\usepackage{amsfonts}
\usepackage{amssymb}

\begin{document}

\author{Antony Valentini\\Augustus College}

\begin{center}
{\LARGE Signal-Locality and Subquantum Information in Deterministic
Hidden-Variables Theories\footnote{To appear in: \textit{Modality,
Probability, and Bell's Theorems}, eds. T. Placek and J. Butterfield (Kluwer,
2002).}}

\bigskip

\bigskip\bigskip

\bigskip Antony Valentini\footnote{email: a.valentini@ic.ac.uk}

\bigskip

\bigskip

\textit{Theoretical Physics Group, Blackett Laboratory, Imperial College,
Prince Consort Road, London SW7 2BZ, England.\footnote{Corresponding address.}}

\textit{Center for Gravitational Physics and Geometry, Department of Physics,
The Pennsylvania State University, University Park, PA 16802, USA.}

\textit{Augustus College, 14 Augustus Road, London SW19 6LN,
England.\footnote{Permanent address.}}

\bigskip
\end{center}

\bigskip

\bigskip

\bigskip

It is proven that any deterministic hidden-variables theory, that reproduces
quantum theory for a `quantum equilibrium' distribution of hidden variables,
must predict the existence of instantaneous signals at the statistical level
for hypothetical `nonequilibrium ensembles'. This `signal-locality theorem'
generalises yet another feature of the pilot-wave theory of de Broglie and
Bohm, for which it is already known that signal-locality is true only in
equilibrium. Assuming certain symmetries, lower bounds are derived on the
`degree of nonlocality' of the singlet state, defined as the (equilibrium)
fraction of outcomes at one wing of an EPR-experiment that change in response
to a shift in the distant angular setting. It is shown by explicit calculation
that these bounds are satisfied by pilot-wave theory. The degree of
nonlocality is interpreted as the average number of bits of `subquantum
information' transmitted superluminally, for an equilibrium ensemble. It is
proposed that this quantity might provide a novel measure of the entanglement
of a quantum state, and that the field of quantum information would benefit
from a more explicit hidden-variables approach. It is argued that the
signal-locality theorem supports the hypothesis, made elsewhere, that in the
remote past the universe relaxed to a state of statistical equilibrium at the
hidden-variable level, a state in which nonlocality happens to be masked by
quantum noise.

\bigskip\bigskip

\bigskip

\bigskip

\bigskip

\bigskip

\bigskip

\bigskip

\bigskip

\bigskip

\bigskip

\bigskip

\bigskip

\bigskip

\bigskip

\bigskip

\bigskip

\bigskip

\bigskip

\bigskip

\bigskip

\bigskip

\bigskip

\bigskip

\bigskip

\bigskip

\bigskip

\bigskip

\bigskip

\section{\bigskip Introduction}

Bell's theorem shows that, with reasonable assumptions, any deterministic
hidden-variables theory behind quantum mechanics has to be nonlocal
[1].\footnote{It is assumed in particular that there is no `conspiracy' or
common cause between the hidden variables and the measurement settings, and
that there is no backwards causation (so that the hidden variables are
unaffected by the future outcomes). Bell's original paper addressed only the
deterministic case. The later generalisations to stochastic theories are of no
concern here.} Specifically, for pairs of spin-1/2 particles in the singlet
state, the outcomes of spin measurements at one wing must depend
instantaneously on the axis of measurement at the other, distant wing.
Historically, Bell's theorem was inspired by a specific nonlocal
hidden-variables theory: the pilot-wave theory of de Broglie and Bohm
[2--10].\footnote{Note that, contrary to a widespread misunderstanding, at the
1927 Fifth Solvay Congress de Broglie proposed the full pilot-wave dynamics in
configuration space for a many-body system, and not just the one-body theory.
See ref. [2].} In his famous review article [11], Bell asked if \textit{all}
hidden-variables theories that reproduce the quantum distribution of outcomes
have to be nonlocal like pilot-wave theory. He subsequently proved that this
is indeed the case.

A further property of pilot-wave theory was also proved to be a general
feature, namely `contextuality'. In general, so-called quantum measurements
are not faithful: they do not reveal the value of an attribute of the system
existing prior to the `measurement'. This property of pilot-wave theory was
discovered by Bohm [3], and the Kochen-Specker theorem [12] tells us that any
hidden-variables interpretation of quantum mechanics must share this
property.\footnote{Even if the original paper by Kochen and Specker
erroneously claimed to prove the nonexistence of hidden variables.}

The question naturally arises: are there any other properties of pilot-wave
theory that are actually universal, in the sense of necessarily being
properties of any viable hidden-variables theory? In this paper we shall prove
that, indeed, yet another feature of pilot-wave theory -- the `signal-locality
theorem' -- is in fact generally true in any deterministic hidden-variables
interpretation: there are instantaneous signals at the statistical level for
hypothetical `nonequilibrium' ensembles whose distribution differs from that
of quantum theory.

We shall also obtain -- assuming certain symmetries -- lower bounds on the
nonlocal flow of `subquantum information' between entangled systems, and we
shall check in detail that these bounds are satisfied (indeed saturated) by
pilot-wave theory. It will be suggested that the `degree of nonlocality',
which quantifies nonlocal information flow, be explored as a new measure of
entanglement, and that the field of quantum information generally would
benefit from a more explicit hidden-variables perspective.

Finally, it will be urged that the results of this paper be viewed in a
cosmological context, as supporting the hypothesis that quantum theory is
merely a theory of an equilibrium state, to which the universe relaxed in the
remote past, perhaps soon after the big bang.

\section{Signal-Locality in Pilot-Wave Theory}

In pilot-wave theory, an individual system with wavefunction $\psi(x,t)$ is
assumed to have a definite configuration $x(t)$ at all times, whose velocity
is determined by the de Broglie guidance equation $\dot{x}(t)=j(x,t)/|\psi
(x,t)|^{2}$ where $j$ is the usual quantum probability current. Given
$\psi(x,t)$, this formula determines the velocity $\dot{x}(t)$. The
wavefunction $\psi$ is interpreted as an objective `guiding field' in
configuration space, and satisfies the usual Schr\"{o}dinger equation.

To recover quantum theory, it must also be assumed that an ensemble of systems
with wavefunction $\psi_{0}(x)$ at $t=0$ begins with a distribution of
configurations given by $\rho_{0}(x)=|\psi_{0}(x)|^{2}$ (which guarantees that
$\rho(x,t)=|\psi(x,t)|^{2}$ at all future times). In other words, the Born
probability distribution is assumed as an initial condition. In principle,
however, the theory -- considered as a theory of dynamics -- allows one to
consider arbitrary initial distributions $\rho_{0}(x)\neq|\psi_{0}(x)|^{2}$,
which violate quantum theory. The `quantum equilibrium' distribution
$\rho=|\psi|^{2}$ is analogous to thermal equilibrium in classical mechanics:
in both cases, the underlying dynamical theory allows one to consider
nonequilibrium; and in both cases equilibrium may be accounted for on the
basis of an appropriate \textit{H}-theorem [13, 14]. Thus, pilot-wave theory
is richer than quantum theory, containing the latter as a special case. (It
has in fact been argued that nonequilibrium $\rho\neq|\psi|^{2}$ existed in
the early universe, and may still exist today for some relic cosmological
particles [5, 10, 13--17].)

Now at the fundamental hidden-variable level, pilot-wave theory is nonlocal.
For example, for two entangled particles $A$ and $B$ the wavefunction
$\psi(x_{A},x_{B},t)$ has a non-separable phase $S(x_{A},x_{B},t)$ and the
velocity $dx_{A}/dt=\nabla_{A}S(x_{A},x_{B},t)/m$ of particle $A$ depends
instantaneously on the position $x_{B}$ of particle $B$. And in general,
operations performed at $B$ (such as switching on an external potential) have
an instantaneous effect on the motion of particle $A$ no matter how distant it
may be.

However, at the quantum level, where one considers an ensemble with the
equilibrium distribution $\rho(x_{A},x_{B},t)=|\psi(x_{A},x_{B},t)|^{2}$,
operations at $B$ have no statistical effect at $A$: as is well known, quantum
entanglement cannot be used for signalling at a distance.

On the other hand, this masking of nonlocality by statistical noise is
peculiar to the distribution $\rho=|\psi|^{2}$. If one considers an ensemble
of entangled particles at $t=0$ with distribution $\rho_{0}(x_{A},x_{B}%
)\neq|\psi_{0}(x_{A},x_{B})|^{2}$, it may be shown by explicit calculation
that changing the Hamiltonian at $B$ induces an instantaneous change in the
marginal distribution $\rho_{A}(x_{A},t)\equiv%
{\textstyle\int}
dx_{B}\ \rho(x_{A},x_{B},t)$ at $A$. For a specific example it was found that
a sudden change $\hat{H}_{B}\rightarrow\hat{H}_{B}^{\prime}$ at $B$ -- say a
change in potential -- leads after a short time $\epsilon$ to a change
$\Delta\rho_{A}\equiv\rho_{A}(x_{A},\epsilon)-\rho_{A}(x_{A},0)$ at $A$ given
by [15]%
\[
\Delta\rho_{A}=-\frac{\epsilon^{2}}{4m}\frac{\partial}{\partial x_{A}}\left(
a(x_{A})\int dx_{B}\ b(x_{B})\frac{\rho_{0}(x_{A},x_{B})-|\psi_{0}(x_{A}%
,x_{B})|^{2}}{|\psi_{0}(x_{A},x_{B})|^{2}}\right)
\]
(Here $a(x_{A})$ is a factor depending on $\psi_{0}$; the factor $b(x_{B})$
also depends on $\hat{H}_{B}^{\prime}$ and vanishes if $\hat{H}_{B}^{\prime
}=\hat{H}_{B}$.) In equilibrium $\rho_{0}=|\psi_{0}|^{2}$ the signal vanishes;
while in general, for $\rho_{0}\neq|\psi_{0}|^{2}$ there are instantaneous
signals at the statistical level.\footnote{Of course, the signal may vanish
for some special $\rho_{0}\neq|\psi_{0}|^{2}$, but not in general.}

This is the signal-locality theorem of pilot-wave theory: in general, there
are instantaneous signals at the statistical level if and only if the ensemble
is in quantum nonequilibrium $\rho_{0}\neq|\psi_{0}|^{2}$ [15]. We wish to
show that the same is true in any deterministic hidden-variables theory.

\section{Bell Nonlocality}

It is convenient first of all to review Bell's theorem in its original
formulation [1].

Consider two spin-1/2 particles lying on the $y$-axis at $A$ and $B$ and
separated by a large distance. If the pair is in the singlet state $\left|
\Psi\right\rangle =\left(  \left|  z+,z-\right\rangle -\left|
z-,z+\right\rangle \right)  /\surd2$, spin measurements along the $z$-axis at
each wing always yield opposite results. But we are of course free to measure
spin components along arbitrary axes at each wing. For simplicity we take the
measurement axes to lie in the $x-z$ plane, so that their orientations may be
specified by the (positive or negative) angles $\theta_{A},$ $\theta_{B}$ made
with the $z$-axis. In units of $\hslash/2$, the possible values of outcomes of
spin measurements along $\theta_{A},$ $\theta_{B}$ at $A,$ $B$ -- that is, the
possible values of the quantum `observables' $\hat{\sigma}_{A},$ $\hat{\sigma
}_{B}$ at $A,$ $B$ -- are $\pm1$. Quantum theory predicts that for an ensemble
of such pairs, the outcomes at $A$ and $B$ are correlated: $\left\langle
\Psi\right|  \hat{\sigma}_{A}\hat{\sigma}_{B}\left|  \Psi\right\rangle
=-\cos(\theta_{A}-\theta_{B})$.

One now assumes the existence of hidden variables $\lambda$ that determine the
outcomes $\sigma_{A},$ $\sigma_{B}=\pm1$ along $\theta_{A},$ $\theta_{B}$. It
is further assumed that there exists a `quantum equilibrium ensemble' of
$\lambda$ -- that is, a distribution $\rho_{eq}(\lambda)$ that reproduces the
quantum statistics (where $\int d\lambda\ \rho_{eq}(\lambda)=1$). Each value
of $\lambda$ determines a pair of outcomes $\sigma_{A},$ $\sigma_{B}$ (for
given $\theta_{A},$ $\theta_{B})$; for an ensemble of similar experiments --
in which the values of $\lambda$ generally differ from one run to the next --
one obtains a distribution of $\sigma_{A},$ $\sigma_{B}$, which is assumed to
agree with quantum theory. In particular, the expectation value%
\[
\overline{\sigma_{A}\sigma_{B}}=\int d\lambda\ \rho_{eq}(\lambda)\sigma
_{A}(\theta_{A},\theta_{B},\lambda)\sigma_{B}(\theta_{A},\theta_{B},\lambda)
\]
must reproduce the quantum result $\left\langle \hat{\sigma}_{A}\hat{\sigma
}_{B}\right\rangle =-\cos(\theta_{A}-\theta_{B})$. Bell showed that this is
possible only if one has nonlocal equations%
\[
\sigma_{A}=\sigma_{A}(\theta_{A},\theta_{B},\lambda),\;\;\;\;\;\sigma
_{B}=\sigma_{B}(\theta_{A},\theta_{B},\lambda)
\]
in which the outcomes depend on the distant angular settings
[1].\footnote{Here $\lambda$ are the \textit{initial} values of the hidden
variables, for example just after the source has produced the singlet pair.
Their later values may be affected by changes in $\theta_{A},$ $\theta_{B}$,
and writing $\sigma_{A}=\sigma_{A}(\theta_{A},\theta_{B},\lambda),$
$\sigma_{B}=\sigma_{B}(\theta_{A},\theta_{B},\lambda)$ (where $\lambda$ are
initial values) allows for this. See ref. [4], chapter 8.}

In principle, the nonlocality might be just `one-way', with only one of
$\sigma_{A},$ $\sigma_{B}$ depending on the distant setting. For instance, one
might have $\sigma_{A}=\sigma_{A}(\theta_{A},\theta_{B},\lambda)$ but
$\sigma_{B}=\sigma_{B}(\theta_{B},\lambda)$, with nonlocality from $B$ to $A$
but not from $A$ to $B$.

\section{General Signal-Locality Theorem}

Consider, then, a nonlocal and deterministic hidden-variables theory that
reproduces quantum theory for some `equilibrium' distribution $\rho
_{eq}(\lambda)$ of hidden variables. (Such theories certainly exist, and
pilot-wave theory provides an example.)

Given a distribution $\rho_{eq}(\lambda)$, one can always contemplate --
purely theoretically -- a `nonequilibrium' distribution $\rho(\lambda)\neq
\rho_{eq}(\lambda)$, even if one cannot prepare such a distribution in
practice. For example, given an ensemble of values of $\lambda$ with
distribution $\rho_{eq}(\lambda)$, mathematically one could pick a subensemble
such that $\rho(\lambda)\neq\rho_{eq}(\lambda)$.

The theorem to be proved is then the following: in general, there are
instantaneous signals at the statistical level if and only if the ensemble is
in quantum nonequilibrium $\rho(\lambda)\neq\rho_{eq}(\lambda)$.

\textit{Proof}: Assume first that $\sigma_{A}$ has some dependence on the
distant setting $\theta_{B}$. (Bell's theorem requires some nonlocal
dependence in at least one direction.)

Now consider an ensemble of experiments with fixed settings $\theta_{A}%
,\theta_{B}$ and an equilibrium distribution $\rho_{eq}(\lambda)$ of hidden
variables $\lambda$. In each experiment, a particular value of $\lambda$
determines an outcome $\sigma_{A}=\sigma_{A}(\theta_{A},\theta_{B},\lambda)$
at $A$. Some values of $\lambda$ yield $\sigma_{A}=+1$, some yield $\sigma
_{A}=-1$. What happens if the setting $\theta_{B}$ at $B$ is changed to
$\theta_{B}^{\prime}$?

The set $S=\left\{  \lambda\right\}  $ of possible values of $\lambda$ may be
partitioned in two ways:%
\[
S_{A+}=\left\{  \lambda|\sigma_{A}(\theta_{A},\theta_{B},\lambda)=+1\right\}
,\;\;\ S_{A-}=\left\{  \lambda|\sigma_{A}(\theta_{A},\theta_{B},\lambda
)=-1\right\}
\]
where $S=S_{A+}\cup S_{A-}$, $S_{A+}\cap S_{A-}=\emptyset$, and%
\[
S_{A+}^{\prime}=\left\{  \lambda|\sigma_{A}(\theta_{A},\theta_{B}^{\prime
},\lambda)=+1\right\}  ,\;\;\ S_{A-}^{\prime}=\left\{  \lambda|\sigma
_{A}(\theta_{A},\theta_{B}^{\prime},\lambda)=-1\right\}
\]
where $S=S_{A+}^{\prime}\cup S_{A-}^{\prime}$, $S_{A+}^{\prime}\cap
S_{A-}^{\prime}=\emptyset$. (There could exist a pathological subset of $S$
that gives neither outcome $\sigma_{A}=\pm1$, but this must have measure zero
with respect to the equilibrium measure $\rho_{eq}(\lambda)$, and so may be
ignored.) It cannot be the case that $S_{A+}=S_{A+}^{\prime}$ and
$S_{A-}=S_{A-}^{\prime}$ for arbitrary $\theta_{B}^{\prime}$, for otherwise
the outcomes at $A$ would not depend at all on the distant setting at $B$.
Thus in general%
\[
T_{A}(+,-)\equiv S_{A+}\cap S_{A-}^{\prime}\neq\emptyset,\;\ \;\;\;T_{A}%
(-,+)\equiv S_{A-}\cap S_{A+}^{\prime}\neq\emptyset
\]
In other words: under a shift $\theta_{B}\rightarrow\theta_{B}^{\prime}$ in
the setting at $B$, some values of $\lambda$ that would have yielded the
outcome $\sigma_{A}=+1$ at $A$ now yield $\sigma_{A}=-1$; and some $\lambda$
that would have yielded $\sigma_{A}=-1$ now yield $\sigma_{A}=+1$.

Of the equilibrium ensemble with distribution $\rho_{eq}(\lambda)$, a fraction%
\[
\nu_{A}^{eq}(+,-)=\int\nolimits_{T_{A}(+,-)}d\lambda\ \rho_{eq}(\lambda)
\]
make the nonlocal `transition' $\sigma_{A}=+1\rightarrow\sigma_{A}=-1$ under
the distant shift $\theta_{B}\rightarrow\theta_{B}^{\prime}$. Similarly, a
fraction%
\[
\nu_{A}^{eq}(-,+)=\int\nolimits_{T_{A}(-,+)}d\lambda\ \rho_{eq}(\lambda)
\]
make the `transition' $\sigma_{A}=-1\rightarrow\sigma_{A}=+1$ under
$\theta_{B}\rightarrow\theta_{B}^{\prime}$.

Now with the initial setting $\theta_{A},\theta_{B}$, quantum theory tells us
that one half of the equilibrium ensemble of values of $\lambda$ yield
$\sigma_{A}=+1$ and the other half yield $\sigma_{A}=-1$. (That is, the
equilibrium measures of $S_{A+}$ and $S_{A-}$ are both $1/2$.) With the new
setting $\theta_{A},\theta_{B}^{\prime}$, quantum theory again tells us that
one half yield $\sigma_{A}=+1$ and the other half yield $\sigma_{A}=-1$ (the
equilibrium measures of $S_{A+}^{\prime}$ and $S_{A-}^{\prime}$ again being
$1/2$). The 1:1 ratio of outcomes $\sigma_{A}=\pm1$ is preserved under the
shift $\theta_{B}\rightarrow\theta_{B}^{\prime}$, from which we deduce the
condition of `detailed balancing'%
\[
\nu_{A}^{eq}(+,-)=\nu_{A}^{eq}(-,+)
\]
The fraction of the equilibrium ensemble that makes the transition $\sigma
_{A}=+1\rightarrow\sigma_{A}=-1$ must equal the fraction that makes the
reverse transition $\sigma_{A}=-1\rightarrow\sigma_{A}=+1$.

But for an arbitrary \textit{non}equilibrium ensemble with distribution
$\rho(\lambda)\neq\rho_{eq}(\lambda)$, the `transition sets' $T_{A}(+,-)$ and
$T_{A}(-,+)$ will generally have different measures%
\[
\int\nolimits_{T_{A}(+,-)}d\lambda\ \rho(\lambda)\neq\int\nolimits_{T_{A}%
(-,+)}d\lambda\ \rho(\lambda)
\]
and the nonequilibrium transition fractions will generally be unequal,%
\[
\nu_{A}(+,-)\neq\nu_{A}(-,+)
\]
(Note that $T_{A}(+,-)$ and $T_{A}(-,+)$ are fixed by the underlying
deterministic theory, and are therefore independent of $\rho(\lambda)$.) Thus,
if with the initial setting $\theta_{A},\theta_{B}$ we would have obtained a
certain nonequilibrium ratio of outcomes $\sigma_{A}=\pm1$ at $A$, with the
new setting $\theta_{A},\theta_{B}^{\prime}$ we will obtain a
\textit{different} ratio at $A$. Under a shift $\theta_{B}\rightarrow
\theta_{B}^{\prime}$, the number of systems that change from $\sigma_{A}=+1$
to $\sigma_{A}=-1$ is unequal to the number that change from $\sigma_{A}=-1$
to $\sigma_{A}=+1$, causing an imbalance that changes the outcome ratios at
$A$. In other words, in general the statistical distribution of outcomes at
$A$ is altered by the distant shift $\theta_{B}\rightarrow\theta_{B}^{\prime}%
$, and there is a statistical signal from $B$ to $A$. (Of course, the signal
vanishes for special $\rho(\lambda)\neq\rho_{eq}(\lambda)$ that happen to have
equal measures for $T_{A}(+,-)$ and $T_{A}(-,+)$, but not in general.)

Similarly, if $\sigma_{B}$ depends on the distant setting $\theta_{A}$, one
may define non-zero transition sets $T_{B}(+,-)$ and $T_{B}(-,+)$ `from $A$ to
$B$'; and in nonequilibrium there will generally be statistical signals from
$A$ to $B$.

In the special case of `one-way' nonlocality, only one of the pairs
$T_{A}(+,-)$, $T_{A}(-,+)$ or $T_{B}(+,-)$, $T_{B}(-,+)$ has non-zero measure,
and nonequilibrium signalling occurs in one direction only.

As an illustrative example of signalling from $B$ to $A$, one might have a
theory in which the variables $\lambda$ consist of pairs of real numbers
$(p,q)$ confined to the area of a unit circle centred on the origin
($p^{2}+q^{2}\leq1$). Imagine that, for the initial setting $\theta_{A}%
,\theta_{B}$, the right half of the circle yields $\sigma_{A}=+1$ while the
left half yields $\sigma_{A}=-1$ (that is, $S_{A+}=\left\{
(p,q)|\ p>0\right\}  $, $S_{A-}=\left\{  (p,q)|\ p<0\right\}  $). Imagine
further that under a small shift $\theta_{B}\rightarrow\theta_{B}^{\prime}$
the vertical chord dividing the circle into $S_{A+}$ and $S_{A-}$ rotates
slightly about the origin, with the area to the right of the rotated chord
yielding $\sigma_{A}=+1$ and the area to the left yielding $\sigma_{A}=-1$. If
we take $\rho_{eq}(\lambda)$ to be uniformly distributed over the area of the
circle, we obtain the quantum 1:1 ratio of outcomes $\sigma_{A}=\pm1$, both
before and after the shift $\theta_{B}\rightarrow\theta_{B}^{\prime}$. But in
general, for a nonuniform distribution $\rho(\lambda)\neq\rho_{eq}(\lambda)$,
not only will the outcome ratio at $A$ with the initial setting $\theta
_{A},\theta_{B}$ be different from 1:1, the ratio at $A$ will \textit{change}
as the chord dividing the circle is rotated by the shift $\theta
_{B}\rightarrow\theta_{B}^{\prime}$.

We have said that, even if $\sigma_{A}$ depends on $\theta_{B}$, the signal
from $B$ to $A$ will vanish for special $\rho(\lambda)\neq\rho_{eq}(\lambda)$
such that the transition sets $T_{A}(+,-)$, $T_{A}(-,+)$ have equal measure.
This hardly affects our argument: the point remains that for a general
nonequilibrium distribution the measures will be unequal and there will be a
signal. At the same time, it would be interesting to know if the signal can
vanish for some special $\rho(\lambda)\neq\rho_{eq}(\lambda)$ for all angular
settings or only for some. (Clearly, for specific angles $\theta_{A}%
,\theta_{B},\theta_{B}^{\prime}$ and associated sets $T_{A}(+,-)$,
$T_{A}(-,+)$, one may trivially choose a special $\rho(\lambda)\neq\rho
_{eq}(\lambda)$ such that $T_{A}(+,-)$, $T_{A}(-,+)$ have the same measure --
for example, $\rho(\lambda)\varpropto\rho_{eq}(\lambda)$ for $\lambda\in
T_{A}(+,-)\cup T_{A}(-,+)$ and $\rho(\lambda)=0$ otherwise. But it is not
known whether there can exist a $\rho(\lambda)\neq\rho_{eq}(\lambda)$ that,
like $\rho_{eq}(\lambda)$, has equal measures for $T_{A}(+,-)$, $T_{A}(-,+)$
for \textit{all} $\theta_{A},\theta_{B},\theta_{B}^{\prime}$. One suspects
not, but the matter should be examined further.)

\section{Degree of Nonlocality}

The possibility of nonlocal signalling from $B$ to $A$ (or from $A$ to $B$)
depends on the existence of finite transition sets $T_{A}(+,-)$, $T_{A}(-,+)$
(or $T_{B}(+,-)$, $T_{B}(-,+)$). We have shown by a detailed-balancing
argument that $T_{A}(+,-)$, $T_{A}(-,+)$ (or $T_{B}(+,-)$, $T_{B}(-,+)$) have
equal equilibrium measure, so that the signal vanishes in equilibrium
$\rho(\lambda)=\rho_{eq}(\lambda)$. On the other hand, a general
nonequilibrium distribution $\rho(\lambda)\neq\rho_{eq}(\lambda)$ will imply
different measures for $T_{A}(+,-)$, $T_{A}(-,+)$ (or $T_{B}(+,-)$,
$T_{B}(-,+)$), resulting in signalling from $B$ to $A$ (or from $A$ to $B$).

But how large can the signal be? There is no signal at all in equilibrium
$\rho(\lambda)=\rho_{eq}(\lambda)$; while if $\rho(\lambda)$ is concentrated
on just one of $T_{A}(+,-)$, $T_{A}(-,+)$ (or on just one of $T_{B}(+,-)$,
$T_{B}(-,+)$), then all the outcomes are changed by the distant shift. Thus
the size of the signal -- measured by the fraction of outcomes that change at
a distance -- can range from 0\% to 100\%.

Now Bell's theorem guarantees that at least one of the pairs $T_{A}(+,-)$,
$T_{A}(-,+)$ or $T_{B}(+,-)$, $T_{B}(-,+)$ has non-zero equilibrium measure:
otherwise we would have a local theory. But so far we have no idea how large
these sets have to be; we know only that $T_{A}(+,-)$ and $T_{A}(-,+)$ must
have equal equilibrium measure, as must $T_{B}(+,-)$ and $T_{B}(-,+)$. (In
this sense, Bell's theorem tells us there must be some nonlocality hidden
behind the equilibrium distribution, but not how much.) The size of the
transition sets is important because if they have very tiny equilibrium
measure, then to obtain an appreciable signal the nonequilibrium distribution
$\rho(\lambda)\neq\rho_{eq}(\lambda)$ would have to be very far from
equilibrium -- that is, concentrated on a very tiny (with respect to the
equilibrium measure) set. We shall therefore try to deduce the equilibrium
measure of the transition sets.

In other words, we now ask the following quantitative question: for an
equilibrium distribution $\rho_{eq}(\lambda)$ of hidden variables, what
fraction of outcomes at $A$ are changed by the distant shift $\theta
_{B}\rightarrow\theta_{B}^{\prime}$, and what fraction at $B$ are changed by
$\theta_{A}\rightarrow\theta_{A}^{\prime}$?

The quantity%
\[
\alpha\equiv\nu_{A}^{eq}(+,-)+\nu_{A}^{eq}(-,+)
\]
(the sum of the equilibrium measures of $T_{A}(+,-)$ and $T_{A}(-,+)$) is the
fraction of the equilibrium ensemble for which the outcomes at $A$ are changed
under $\theta_{B}\rightarrow\theta_{B}^{\prime}$ (irrespective of whether they
change from $+1$ to $-1$ or vice versa, the fractions doing each being
$\alpha/2$). There is a `degree of nonlocality from $B$ to $A$', quantified by
$\alpha=\alpha(\theta_{A},\theta_{B},\theta_{B}^{\prime})$. Similarly, one may
define a `degree of nonlocality from $A$ to $B$', quantified by the fraction
$\beta=\beta(\theta_{A},\theta_{B},\theta_{A}^{\prime})$ of outcomes at $B$
that change in response to a shift $\theta_{A}\rightarrow\theta_{A}^{\prime}$
at $A$. Bell's theorem tells us that, in general, the `total degree of
nonlocality' $\alpha+\beta>0$.\footnote{Of course, $\alpha+\beta$ could vanish
for specific angles, but not in general.} Positive lower bounds on
$\alpha+\beta$, and on $\alpha$ or $\beta$ alone, may be obtained if one
assumes certain symmetries.

\section{A General Lower Bound}

First, we derive a general lower bound for the quantity $\alpha+\tilde{\beta}%
$, where%
\[
\tilde{\beta}\equiv\nu_{B}^{eq}(-,+)+\nu_{B}^{eq}(+,-)
\]
is the equilibrium fraction of outcomes that change at $B$, under the
\textit{local} shift $\theta_{B}\rightarrow\theta_{B}^{\prime}$ (with $\nu
_{B}^{eq}(-,+)$ and $\nu_{B}^{eq}(+,-)$ defined similarly to $\nu_{A}%
^{eq}(-,+)$ and $\nu_{A}^{eq}(+,-)$ above). In other words, we obtain a lower
bound on the sum of the nonlocal and local effects of $\theta_{B}%
\rightarrow\theta_{B}^{\prime}$.

The quantity $\frac{1}{2}\left|  \sigma_{A}(\theta_{A},\theta_{B}^{\prime
},\lambda)-\sigma_{A}(\theta_{A},\theta_{B},\lambda)\right|  $ equals $1$ if
the outcome $\sigma_{A}$ changes under $\theta_{B}\rightarrow\theta
_{B}^{\prime}$, and vanishes otherwise. Since $\rho_{eq}(\lambda)\,d\lambda$
is by definition the fraction of the equilibrium ensemble for which $\lambda$
lies in the interval $(\lambda,\lambda+d\lambda)$, the fraction for which
$\sigma_{A}$ changes is%
\[
\alpha=\frac{1}{2}\int d\lambda\ \rho_{eq}(\lambda)\left|  \sigma_{A}%
(\theta_{A},\theta_{B}^{\prime},\lambda)-\sigma_{A}(\theta_{A},\theta
_{B},\lambda)\right|
\]
Similarly, the fraction for which $\sigma_{B}$ changes is%
\[
\tilde{\beta}=\frac{1}{2}\int d\lambda\ \rho_{eq}(\lambda)\left|  \sigma
_{B}(\theta_{A},\theta_{B}^{\prime},\lambda)-\sigma_{B}(\theta_{A},\theta
_{B},\lambda)\right|
\]
Now%
\[
-\cos(\theta_{A}-\theta_{B})=\int d\lambda\ \rho_{eq}(\lambda)\sigma
_{A}(\theta_{A},\theta_{B},\lambda)\sigma_{B}(\theta_{A},\theta_{B},\lambda)
\]
and so%
\[
\left|  \cos(\theta_{A}-\theta_{B}^{\prime})-\cos(\theta_{A}-\theta
_{B})\right|  \leq
\]%
\[
\int d\lambda\ \rho_{eq}(\lambda)\left|  \sigma_{A}(\theta_{A},\theta
_{B}^{\prime},\lambda)\sigma_{B}(\theta_{A},\theta_{B}^{\prime},\lambda
)-\sigma_{A}(\theta_{A},\theta_{B},\lambda)\sigma_{B}(\theta_{A},\theta
_{B},\lambda)\right|
\]%
\[
=\int d\lambda\ \rho_{eq}(\lambda)\left|
\begin{array}
[c]{c}%
\sigma_{A}(\theta_{A},\theta_{B}^{\prime},\lambda)\sigma_{B}(\theta_{A}%
,\theta_{B}^{\prime},\lambda)-\sigma_{A}(\theta_{A},\theta_{B},\lambda
)\sigma_{B}(\theta_{A},\theta_{B}^{\prime},\lambda)\\
+\sigma_{A}(\theta_{A},\theta_{B},\lambda)\sigma_{B}(\theta_{A},\theta
_{B}^{\prime},\lambda)-\sigma_{A}(\theta_{A},\theta_{B},\lambda)\sigma
_{B}(\theta_{A},\theta_{B},\lambda)
\end{array}
\right|
\]%
\begin{align*}
&  \leq\int d\lambda\ \rho_{eq}(\lambda)\left|  \sigma_{A}(\theta_{A}%
,\theta_{B}^{\prime},\lambda)-\sigma_{A}(\theta_{A},\theta_{B},\lambda)\right|
\\
&  +\int d\lambda\ \rho_{eq}(\lambda)\left|  \sigma_{B}(\theta_{A},\theta
_{B}^{\prime},\lambda)-\sigma_{B}(\theta_{A},\theta_{B},\lambda)\right|
\end{align*}%

\[
=2\alpha+2\tilde{\beta}%
\]
Thus we have the lower bound%
\begin{equation}
\alpha(\theta_{A},\theta_{B},\theta_{B}^{\prime})+\tilde{\beta}(\theta
_{A},\theta_{B},\theta_{B}^{\prime})\geq\frac{1}{2}\left|  \cos(\theta
_{A}-\theta_{B}^{\prime})-\cos(\theta_{A}-\theta_{B})\right|
\end{equation}

The maximum value of the right hand side is $1$. From this inequality alone,
then, one could have $\alpha$ arbitrarily close to zero, with $\tilde{\beta
}\rightarrow1$ -- that is, an arbitrarily small fraction could change at $A$
in response to $\theta_{B}\rightarrow\theta_{B}^{\prime}$, provided virtually
all the outcomes change at $B$. So far, then, we have no lower bound on the
nonlocal effect from $B$ to $A$, as quantified by $\alpha$: it is only the sum
$\alpha+\tilde{\beta}$ of the nonlocal and local effects that is bounded.

\section{Symmetric Cases}

A lower bound on $\alpha+\beta$ -- the sum of the nonlocal effects from $B$ to
$A$ and from $A$ to $B$ -- may be obtained if we assume an appropriate
rotational symmetry at the hidden-variable level. A lower bound on $\alpha$ or
$\beta$ alone may be obtained if we also assume that the measurement
operations at $A$ and $B$ are identical -- that is, use identical equipment
and coupling -- so that there is symmetry between the two wings.

\textit{Rotational Symmetry}: Consider the effect of a shift $\theta
_{B}\rightarrow\theta_{B}^{\prime}=\theta_{B}+\delta$ at $B$. This changes
certain fractions $\alpha(\theta_{A},\theta_{B},\theta_{B}+\delta)$ and
$\tilde{\beta}(\theta_{A},\theta_{B},\theta_{B}+\delta)$ of the (equilibrium)
outcomes at $A$ and $B$ respectively. Let us assume that the same changes are
effected by the shift $\theta_{A}\rightarrow\theta_{A}^{\prime}=\theta
_{A}-\delta$ at $A$. This means that $\beta(\theta_{A},\theta_{B},\theta
_{A}-\delta)$ -- the fraction of outcomes at $B$ that change (nonlocally) in
response to a shift $\theta_{A}\rightarrow\theta_{A}^{\prime}=\theta
_{A}-\delta$ at $A$ -- is equal to $\tilde{\beta}(\theta_{A},\theta_{B}%
,\theta_{B}+\delta)$, and so from (1)%
\begin{equation}
\alpha(\theta_{A},\theta_{B},\theta_{B}+\delta)+\beta(\theta_{A},\theta
_{B},\theta_{A}-\delta)\geq\frac{1}{2}\left|  \cos(\theta_{A}-\theta
_{B}-\delta)-\cos(\theta_{A}-\theta_{B})\right|
\end{equation}

For $\theta_{A}=\theta_{B}=0$,%
\begin{equation}
\alpha(0,0,\delta)+\beta(0,0,-\delta)\geq\frac{1}{2}(1-\cos\delta)
\end{equation}
Thus, for example, the fraction $\alpha$ that changes at $A$ due to a shift
$+\pi/2$ at $B$ plus the fraction $\beta$ that changes at $B$ due to a shift
$-\pi/2$ at $A$ must be at least 50\%.

Note that our assumption of `rotational symmetry' refers to the
hidden-variable level.\footnote{Quantum theory alone says nothing at all about
$\alpha,\beta,\tilde{\alpha},\tilde{\beta}$. And at the quantum level, the
singlet state is of course rotationally invariant.} How it relates to
fundamental rotational invariance is not clear, for in the absence of a
definite theory one does not know how the hidden variables transform under a
rotation. In any case, the assumption seems stronger than fundamental
rotational invariance: it might be violated if the measuring apparatus happens
to pick out an effectively preferred direction in space.

\textit{Rotational and Exchange Symmetry}: Assuming further an exchange
symmetry between $A$ and $B$ -- specifically, that the effect at $A$ of a
shift $\theta_{B}\rightarrow\theta_{B}^{\prime}=\theta_{B}+\delta$ at $B$
equals the effect at $B$ of a shift $\theta_{A}\rightarrow\theta_{A}^{\prime
}=\theta_{A}-\delta$ at $A$ -- we also have $\alpha(\theta_{A},\theta
_{B},\theta_{B}+\delta)=\beta(\theta_{A},\theta_{B},\theta_{A}-\delta
)$.\footnote{This assumption excludes `one-way' nonlocality.} Thus we obtain a
lower bound%
\begin{equation}
\alpha(\theta_{A},\theta_{B},\theta_{B}+\delta)\geq\frac{1}{4}\left|
\cos(\theta_{A}-\theta_{B}-\delta)-\cos(\theta_{A}-\theta_{B})\right|
\end{equation}
on $\alpha$ (or $\beta$) alone.

For $\theta_{A}=\theta_{B}=0$,%
\begin{equation}
\alpha(0,0,\delta)\geq\frac{1}{4}(1-\cos\delta)
\end{equation}
If the measurement angle at $B$ is shifted by $\pi/2$, at least 25\% of the
outcomes change at $A$ (and similarly from $A$ to $B$).

Clearly, in these symmetric cases, the transition sets are necessarily very
large, and even a mild disequilibrium $\rho(\lambda)\neq\rho_{eq}(\lambda)$
will entail a significant signal.

\section{Comparison with Pilot-Wave Theory}

It behoves us to check that the bounds obtained above are satisfied for the
specific hidden-variables model provided by pilot-wave theory. To do so, we
use the theory of spin measurements due to Bell.\footnote{See ref. [4],
chapter 15.}

\textit{Pilot-Wave Spin Measurements}: At each wing $A$ and $B$ let there be
an apparatus that performs a quantum measurement of spin. The pointer
positions $r_{A}$ and $r_{B}$ begin in `neutral' states centred at the origin.
More precisely, we assume that at $t=0$ the pointers have identical localised
wavepackets $\phi(r_{A})$, $\phi(r_{B})$ centred at $r_{A}$, $r_{B}=0$. During
the measurement, each packet will move `up' or `down' (that is, towards
positive or negative values of $r_{A}$, $r_{B}$), thereby indicating an
outcome of spin up or down.

In the case of a Stern-Gerlach apparatus, the `pointer position' would in fact
be the position of the particle itself as it is deflected upon passing through
the magnetic field. But here, $r_{A}$, $r_{B}$ are simply abstract pointer
positions for measuring equipment that is brought into interaction with the
particles.\footnote{There is in fact an important difference between the
measurement process described here and the Stern-Gerlach method, in the case
of angles differing by $\pi$. This is discussed in section 9 below.}

Consider, then, the initial total wavefunction $\psi_{ij}(r_{A},r_{B}%
,0)=\phi(r_{A})\phi(r_{B})a_{ij}$, where the indices $i,j=\pm$ denote spin up
or down at $A$, $B$ and the $a_{ij}$ are spin amplitudes for the singlet
state. The pointers are initially independent, but the spins are entangled
$a_{ij}\neq b_{i}c_{j}$. The total Hamiltonian%
\[
\hat{H}=g_{A}(t)\hat{\sigma}_{A}\left(  -i\partial/\partial r_{A}\right)
+g_{B}(t)\hat{\sigma}_{B}\left(  -i\partial/\partial r_{B}\right)
\]
(where the couplings $g_{A}(t),$ $g_{B}(t)$ are switched on at $t=0$)
describes ideal von Neumann measurements of spin at $A$ and $B$.\footnote{The
Hamiltonians of the particles and pointers themselves are assumed to be
negligible compared to the interaction part (so that the free spreading of the
pointer packets is negligible). This may be justified for a realistic model by
considering strong couplings over short times; or, one can simply accept the
above as an illustrative model of measurement.} The Schr\"{o}dinger equation
then reads%
\[
i\frac{\partial\psi_{ij}}{\partial t}=g_{A}(t)(\sigma_{A})_{ik}\left(
-i\frac{\partial}{\partial r_{A}}\right)  \psi_{kj}+g_{B}(t)(\sigma_{B}%
)_{jk}\left(  -i\frac{\partial}{\partial r_{B}}\right)  \psi_{ik}%
\]
where repeated indices are summed over, and $\sigma_{A,B}$ is the Pauli spin
matrix $\left(
\begin{array}
[c]{cc}%
1 & 0\\
0 & -1
\end{array}
\right)  $ (where in the spin subspace at $A$ the vector $\left(
\begin{array}
[c]{c}%
1\\
0
\end{array}
\right)  $ denotes spin up along the axis $\theta_{A}$ while in subspace $B$
the vector $\left(
\begin{array}
[c]{c}%
1\\
0
\end{array}
\right)  $ denotes spin up along $\theta_{B}$).

The quantum equilibrium distribution for the pointer positions is%
\[
\rho_{eq}(r_{A},r_{B},t)=\left|  \psi_{++}\right|  ^{2}+\left|  \psi
_{+-}\right|  ^{2}+\left|  \psi_{-+}\right|  ^{2}+\left|  \psi_{--}\right|
^{2}%
\]
From the Schr\"{o}dinger equation one readily derives a continuity equation
for $\rho_{eq}$ and associated probability currents%
\[
j_{A}=g_{A}\psi_{ki}^{\ast}(\sigma_{A})_{kl}\psi_{li}\;,\;\;\;\;j_{B}%
=g_{B}\psi_{ik}^{\ast}(\sigma_{B})_{kl}\psi_{il}%
\]
The hidden-variable pointer positions $r_{A}(t),r_{B}(t)$ have velocities
given by the de Broglie guidance formulas $v_{A,B}=j_{A,B}/\rho_{eq}$.
Explicitly\footnote{Note that the hidden-variable velocities are independent
of the spin basis used.}%
\[
v_{A}=g_{A}\left(  \left|  \psi_{++}\right|  ^{2}+\left|  \psi_{+-}\right|
^{2}-\left|  \psi_{-+}\right|  ^{2}-\left|  \psi_{--}\right|  ^{2}\right)
/\rho_{eq}%
\]%
\[
v_{B}=g_{B}\left(  \left|  \psi_{++}\right|  ^{2}-\left|  \psi_{+-}\right|
^{2}+\left|  \psi_{-+}\right|  ^{2}-\left|  \psi_{--}\right|  ^{2}\right)
/\rho_{eq}%
\]

If $\psi_{ij}(r_{A},r_{B},0)=\phi(r_{A})\phi(r_{B})a_{ij}$ then with%
\[
h_{A}(t)\equiv\int\nolimits_{0}^{t}dt^{\prime}\ g_{A}(t^{\prime}%
),\ \ \ \ \ \ h_{B}(t)\equiv\int\nolimits_{0}^{t}dt^{\prime}\ g_{B}(t^{\prime
})
\]
the Schr\"{o}dinger equation implies that at $t>0$%
\[
\psi_{++}=a_{++}\phi(r_{A}-h_{A})\phi(r_{B}-h_{B}),\ \ \ \ \ \psi_{+-}%
=a_{+-}\phi(r_{A}-h_{A})\phi(r_{B}+h_{B})
\]%
\[
\psi_{-+}=a_{-+}\phi(r_{A}+h_{A})\phi(r_{B}-h_{B}),\ \ \ \ \ \psi_{--}%
=a_{--}\phi(r_{A}+h_{A})\phi(r_{B}+h_{B})
\]
The branches $\psi_{ij}$ eventually separate in configuration space, and the
actual configuration $\left(  r_{A}(t),r_{B}(t)\right)  $ can end up occupying
only one of them -- leading to a definite outcome. For example, if at large
$t$ the actual $\left(  r_{A},r_{B}\right)  $ lies in $\psi_{+-}$ alone, then
the guidance formulas imply $v_{A}=g_{A}$, $v_{B}=-g_{B}$ and the pointer
positions will be $r_{A}(t)\approx h_{A}(t)$, $r_{B}(t)\approx-h_{B}(t)$,
corresponding to $\sigma_{A}=+1$, $\sigma_{B}=-1$.

The outcomes $\sigma_{A}$, $\sigma_{B}$ depend on the hidden variables
$r_{A}(0),r_{B}(0)$ and on the settings $\theta_{A}$, $\theta_{B}$. The
question now is: what fraction of the outcomes change under $\theta
_{B}\rightarrow\theta_{B}^{\prime}$ (or $\theta_{A}\rightarrow\theta
_{A}^{\prime}$)?

More precisely, for given settings $\theta_{A}$, $\theta_{B}$ the outcomes
$\sigma_{A}$, $\sigma_{B}$ are determined by the initial wavefunction
$\psi_{ij}(r_{A},r_{B},0)$ and the initial pointer positions $r_{A}%
(0),r_{B}(0)$, so strictly speaking the `hidden variables' are $\lambda
=\left(  \psi_{ij}(r_{A},r_{B},0),r_{A}(0),r_{B}(0)\right)  $. However,
because $\psi_{ij}(r_{A},r_{B},0)$ is assumed to be the same in every run of
the experiment, there is no need to consider it, and effectively
$\lambda=\left(  r_{A}(0),r_{B}(0)\right)  $. We wish to calculate the
fraction of the equilibrium ensemble of $r_{A}(0),r_{B}(0)$ for which the
outcome $\sigma_{A}$ (or $\sigma_{B}$) changes under $\theta_{B}%
\rightarrow\theta_{B}^{\prime}$ (or $\theta_{A}\rightarrow\theta_{A}^{\prime}$).

If we take square initial pointer packets $\left|  \phi(r_{A})\right|  ^{2}$,
$\left|  \phi(r_{B})\right|  ^{2}$, each equal to $1/\Delta$ for
$-\Delta/2<r_{A},r_{B}<\Delta/2$ and zero elsewhere, then in the $r_{A}-r_{B}$
plane the initial distribution $\rho_{eq}(r_{A},r_{B},0)=\left|  \phi
(r_{A})\right|  ^{2}\left|  \phi(r_{B})\right|  ^{2}$ is uniform within a
square of side $\Delta$ centred on the origin (with no support outside the
square). Fractional areas within the square then represent statistical
fractions of the initial ensemble. Thus, we need to calculate the fractional
area of initial points within the square for which $\sigma_{A}$ (or
$\sigma_{B}$) changes under $\theta_{B}\rightarrow\theta_{B}^{\prime}$ (or
$\theta_{A}\rightarrow\theta_{A}^{\prime}$).

\textit{Symmetric Case}: First we look at a symmetric case with equal
couplings at each wing: $g_{A}(t)=g_{B}(t)=a\,\theta(t)$, where $a$ is a
positive constant, and $h_{A}(t)=h_{B}(t)=at$ (for $t\geq0$). The packets move
with equal speeds along the $r_{A}$- and $r_{B}$-axes.

Taking initial measurement angles $\theta_{A}=\theta_{B}=0$, the singlet state
has spin amplitudes%
\[
a_{++}=0\;,\;\ a_{+-}=1/\surd2\;,\;\ a_{-+}=-1/\surd2\;,\;\ a_{--}=0
\]
We then have $\psi_{++}=\psi_{--}=0$ and the pointer velocities are%
\[
v_{A}=a\frac{\left(  \left|  \psi_{+-}\right|  ^{2}-\left|  \psi_{-+}\right|
^{2}\right)  }{\left(  \left|  \psi_{+-}\right|  ^{2}+\left|  \psi
_{-+}\right|  ^{2}\right)  }\;,\ \ \ \ v_{B}=-v_{A}%
\]
It is then straightforward to deduce that in the $r_{A}-r_{B}$ plane all
initial points $(r_{A}(0),r_{B}(0))$ above the line $r_{B}=r_{A}$ (plotting
$r_{B}$ as ordinate and $r_{A}$ as abscissa) end up in the branch $\psi_{-+}$,
yielding outcomes $\sigma_{A}=-1,$ $\sigma_{B}=+1$; while those below that
line end up in $\psi_{+-}$, yielding $\sigma_{A}=+1,$ $\sigma_{B}=-1$. [Where
$\psi_{+-}$ and $\psi_{-+}$ overlap, $v_{A}=v_{B}=0$; while if $\psi_{+-}%
\neq0$ and $\psi_{-+}=0$ then $v_{A}=+a$, $v_{B}=-a$; and if $\psi_{-+}\neq0$
and $\psi_{+-}=0$ then $v_{A}=-a$, $v_{B}=+a$. It follows that as the branches
separate -- with $\psi_{+-}$ and $\psi_{-+}$ moving perpendicular to the line
$r_{B}=r_{A}$, into the bottom-right and top-left quadrants respectively -- a
point $(r_{A},r_{B})$ initially above the line $r_{B}=r_{A}$ will begin at
rest but will eventually be left behind by the branch $\psi_{+-}$, whereupon
it will acquire the velocity $(v_{A},v_{B})=(-a,+a)$ due to guidance by
$\psi_{-+}$. At large times $r_{A}\approx-at$, $r_{B}\approx+at$ yielding the
outcomes $\sigma_{A}=-1,$ $\sigma_{B}=+1$. Similar reasoning shows that a
point $(r_{A},r_{B})$ initially below the line $r_{B}=r_{A}$ yields the
outcomes $\sigma_{A}=+1,$ $\sigma_{B}=-1$.]

Now, what happens if we change the axis of measurement at $B$ -- say to
$\theta_{B}^{\prime}=\pi/2$? For the settings $\theta_{A}=0,\ \theta
_{B}^{\prime}=\pi/2$, the singlet state has spin amplitudes%
\[
a_{++}^{\prime}=1/2,\ a_{+-}^{\prime}=1/2,\ a_{-+}^{\prime}=-1/2,\ a_{--}%
^{\prime}=1/2
\]
and now all four branches $\psi_{ij}^{\prime}$ contribute to the velocities,
each branch moving into a different quadrant of the $r_{A}-r_{B}$ plane. It is
not difficult to see that a point $(r_{A},r_{B})$ initially in the top-right
quadrant ends up in $\psi_{++}^{\prime}$, yielding $\sigma_{A}=+1,$
$\sigma_{B}=+1$; while points in the bottom-right quadrant yield $\sigma
_{A}=+1,$ $\sigma_{B}=-1$; those in the top-left yield $\sigma_{A}=-1,$
$\sigma_{B}=+1$; and those in the bottom-left yield $\sigma_{A}=-1,$
$\sigma_{B}=-1$. [Where all four branches overlap, both velocity components
vanish, $v_{A}^{\prime}=v_{B}^{\prime}=0$; where just two overlap, one
component vanishes (for example if only $\psi_{++}^{\prime}$ and $\psi
_{+-}^{\prime}$ are non-zero then $v_{A}^{\prime}=+a,\ v_{B}^{\prime}=0$); and
where none overlap neither component vanishes (for example if only $\psi
_{++}^{\prime}$ is non-zero then $v_{A}^{\prime}=+a,\ v_{B}^{\prime}=+a$).
Thus, for example, consider an initial point $(r_{A},r_{B})$ in the top-right
quadrant and below the line $r_{B}=r_{A}$. For as long as all four branches
overlap at $(r_{A},r_{B})$, the point will remain at rest. But after a while
only $\psi_{++}^{\prime}$ and $\psi_{+-}^{\prime}$ will overlap there, and for
an interim period the point will move along $r_{A}$, remaining in the same
quadrant, having acquired velocity components $v_{A}^{\prime}=+a,\ v_{B}%
^{\prime}=0$. Later, once $\psi_{++}^{\prime}$ and $\psi_{+-}^{\prime}$ have
separated, the point is guided by $\psi_{++}^{\prime}$ alone, acquiring
velocity components $v_{A}^{\prime}=+a,\ v_{B}^{\prime}=+a$, and at large
times $r_{A}\approx+at$, $r_{B}\approx+at$, yielding the outcomes $\sigma
_{A}=+1,$ $\sigma_{B}=+1$. For an initial point $(r_{A},r_{B})$ in the
top-right quadrant but above the line $r_{B}=r_{A}$, there is an interim
period of motion along $r_{B}$, again remaining in the same quadrant, after
which the point is again carried by $\psi_{++}^{\prime}$ to $r_{A}\approx+at$,
$r_{B}\approx+at$ at large times, again yielding $\sigma_{A}=+1,$ $\sigma
_{B}=+1$. Thus, all initial points in the top-right quadrant yield $\sigma
_{A}=+1,$ $\sigma_{B}=+1$. Similarly, points in the bottom-right quadrant
yield $\sigma_{A}=+1,$ $\sigma_{B}=-1$, while those in the top-left yield
$\sigma_{A}=-1,$ $\sigma_{B}=+1$ and those in the bottom-left yield
$\sigma_{A}=-1,$ $\sigma_{B}=-1$.]

Clearly, some initial points $(r_{A},r_{B})$ that would have yielded
$\sigma_{A}=+1,$ $\sigma_{B}=-1$ with the settings $\theta_{A}=0,\ \theta
_{B}=0$ still yield $\sigma_{A}=+1,$ $\sigma_{B}=-1$ with $\theta
_{A}=0,\ \theta_{B}^{\prime}=\pi/2$ -- namely, all points in the bottom-right
quadrant. Similarly, all points in the top-left quadrant yield $\sigma
_{A}=-1,$ $\sigma_{B}=+1$ with both the old and new settings. But in the other
two quadrants, the outcomes are changed by the shift from $\theta_{B}=0$ to
$\theta_{B}^{\prime}=\pi/2$. In the top-right, whereas before half gave
$\sigma_{A}=+1,$ $\sigma_{B}=-1$ and half $\sigma_{A}=-1,$ $\sigma_{B}=+1$,
now all give $\sigma_{A}=+1,$ $\sigma_{B}=+1$. In the bottom-left, before half
gave $\sigma_{A}=+1,$ $\sigma_{B}=-1$ and half $\sigma_{A}=-1,$ $\sigma
_{B}=+1$, whereas now all give $\sigma_{A}=-1,$ $\sigma_{B}=-1$.

A simple count shows that 25\% of the outcomes have changed at $A$ and 25\%
have changed at $B$. Thus, under $\theta_{B}=0\rightarrow\theta_{B}^{\prime
}=\pi/2$, the fraction $\alpha(0,0,\pi/2)=1/4$ of outcomes that change at $A$
is indeed equal to the fraction $\tilde{\beta}(0,0,\pi/2)=1/4$ of outcomes
that change at $B$, and our inequality (5) is exactly saturated.

For a shift to an arbitrary angle $\theta_{B}^{\prime}=\delta$ at $B$, we find
that $\alpha(0,0,\delta)=\tilde{\beta}(0,0,\delta)=\frac{1}{4}(1-\cos\delta)$:
again, the fractional changes at $A$ and $B$ are equal, and (5) is exactly
saturated. [With the new settings the singlet state has spin amplitudes%
\begin{align*}
a_{++}^{\prime}  &  =\frac{1}{\sqrt{2}}\sin\frac{\delta}{2}\;,\;\;\;\ a_{+-}%
^{\prime}=\frac{1}{\sqrt{2}}\cos\frac{\delta}{2},\\
\ a_{-+}^{\prime}  &  =-\frac{1}{\sqrt{2}}\cos\frac{\delta}{2}%
\;,\;\;\;\ a_{--}^{\prime}=\frac{1}{\sqrt{2}}\sin\frac{\delta}{2}%
\end{align*}
and again all four branches $\psi_{ij}^{\prime}$ contribute to the velocities
and move into different quadrants of the $r_{A}-r_{B}$ plane. As in the case
$\delta=\pi/2$, where all four branches overlap both velocity components
vanish, $v_{A}^{\prime}=v_{B}^{\prime}=0$; but now, where just two overlap,
neither component vanishes (for example, if only $\psi_{++}^{\prime}$ and
$\psi_{+-}^{\prime}$ are non-zero then $v_{A}^{\prime}=+a,\ v_{B}^{\prime
}=-a\cos\delta$); and where none overlap the velocities are as for $\delta
=\pi/2$ (for example if only $\psi_{++}^{\prime}$ is non-zero then
$v_{A}^{\prime}=+a,\ v_{B}^{\prime}=+a$). Thus, again considering an initial
point $(r_{A},r_{B})$ in the top-right quadrant and below the line
$r_{B}=r_{A}$, for as long as all four branches overlap at $(r_{A},r_{B})$ the
point will remain at rest. As before, after a while only $\psi_{++}^{\prime}$
and $\psi_{+-}^{\prime}$ will overlap there, but now there is an interim
period during which the point not only moves along $r_{A}$ with velocity
$v_{A}^{\prime}=+a$, it also moves along $r_{B}$ with velocity $v_{B}^{\prime
}=-a\cos\delta$ and need \textit{not} remain in the same quadrant. If the
initial point is sufficiently close to the $r_{A}$ axis -- below the line
$r_{B}=(\cos\delta)r_{A}$, in fact -- it may cross that axis into the
bottom-right quadrant before the four branches separate altogether. If that
happens, once the branches separate completely the point will be guided by
$\psi_{+-}^{\prime}$, yielding $\sigma_{A}=+1,$ $\sigma_{B}=-1$. Similarly, an
initial point in the top-right quadrant and above the line $r_{B}=r_{A}$ will
cross into the top-left quadrant -- yielding $\sigma_{A}=-1,$ $\sigma_{B}=+1$
-- if it begins sufficiently close to the $r_{B}$ axis (above the line
$r_{B}=r_{A}/\cos\delta$). Thus, unlike for $\delta=\pi/2$, not every initial
point in the top-right quadrant yields $\sigma_{A}=+1,$ $\sigma_{B}=+1$, but
only those between the lines $r_{B}=(\cos\delta)r_{A}$ and $r_{B}=r_{A}%
/\cos\delta$; the rest cross over into neighbouring quadrants and yield
different outcomes\footnote{We have assumed that $\cos\delta>0$. If
$\cos\delta<0$, instead of points leaving the top-right quadrant they can
enter it from neighbouring quadrants. But the resulting fractional changes at
$A$ and $B$ are the same.}. Similar results are found for the opposite
(bottom-left) quadrant, the fate of initial points being reflection-symmetric
about the line $r_{B}=-r_{A}$. Results for the other two quadrants are as for
$\delta=\pi/2$. Elementary geometry then shows that the regions of the initial
square in which outcomes at $A$ change under $\theta_{B}=0\rightarrow
\theta_{B}^{\prime}=\delta$ have a total fractional area $\frac{1}{4}%
(1-\cos\delta)$, as do regions in which outcomes at $B$ change.]

The maximum change is obtained for $\delta=\pi$, for which 50\% change at $A$
and $B$ ($\alpha(0,0,\pi)=\tilde{\beta}(0,0,\pi)=0.5$). Because our inequality
(5) is exactly saturated for all $\delta$, in a precisely defined sense it may
be said that pilot-wave theory is `minimally nonlocal'. Though whether (5) is
saturated for non-square pointer packets is not known.

\textit{Asymmetric Case}: Next we look at an asymmetric case where the
couplings at each wing are unequal: $g_{A}=a_{A}\,\theta(t)$, $g_{B}%
=a_{B}\,\theta(t)$ with $a_{A}=2a_{B}$ so that the packets move with different
speeds along the $r_{A}$- and $r_{B}$-axes. With the settings $\theta
_{A}=0,\ \theta_{B}=0$ it is found that instead of the $r_{A}-r_{B}$ plane
being divided by the line $r_{B}=r_{A}$ into points yielding $\sigma_{A}=-1,$
$\sigma_{B}=+1$ (above the line) and $\sigma_{A}=+1,$ $\sigma_{B}=-1$ (below
it), as it was for the case $a_{A}=a_{B}$, it is now divided by the line
$r_{B}=\frac{1}{2}r_{A}-\frac{\Delta}{4}$ for $-\frac{\Delta}{2}\leq r_{A}%
\leq0$, the vertical line $r_{A}=0$ for $-\frac{\Delta}{4}\leq r_{B}\leq
+\frac{\Delta}{4}$, and $r_{B}=\frac{1}{2}r_{A}+\frac{\Delta}{4}$ for $0\leq
r_{A}\leq+\frac{\Delta}{2}$. When the angle at $B$ is reset to $\theta
_{B}^{\prime}=\pi/2$, it is found that the four quadrants yield the same
outcomes as above for $a_{A}=a_{B}$ (the top-right yielding $\sigma_{A}=+1,$
$\sigma_{B}=+1$, the bottom-right $\sigma_{A}=+1,$ $\sigma_{B}=-1$, the
top-left $\sigma_{A}=-1,$ $\sigma_{B}=+1$, and the bottom-left $\sigma
_{A}=-1,$ $\sigma_{B}=-1$), there being no crossing from one quadrant into
another. Inspection shows that a fraction $\alpha(0,0,\pi/2)=1/8$ of outcomes
have changed at $A$, while a fraction $\tilde{\beta}(0,0,\pi/2)=3/8$ have
changed at $B$.

Thus, $\alpha\neq\tilde{\beta}$ in this asymmetrical case: the nonlocal and
local effects are unequal, as expected. Further, the general bound (1) is
satisfied, and indeed exactly saturated.

In the limit $a_{A}/a_{B}\rightarrow\infty$ -- where the measurement at $A$
takes place much more rapidly than at $B$ (in the sense of rate of branch
separation) -- it is found that $\alpha(0,0,\pi/2)\rightarrow0$, $\tilde
{\beta}(0,0,\pi/2)\rightarrow1/2$. The bound (1) is again saturated, the
nonlocal effect being arbitrarily small and the local effect approaching 50\%.
[With $\theta_{A}=0,\ \theta_{B}=0$, there are just two branches $\psi_{+-}$
and $\psi_{-+}$ which rapidly separate along $r_{A}$ before hardly any motion
has occurred along $r_{B}$, where in the $r_{A}-r_{B}$ plane $\psi_{+-}$ moves
to the right and $\psi_{-+}$ moves to the left: thus the right half of the
initial square yields $\sigma_{A}=+1,$ $\sigma_{B}=-1$ and the left half
yields $\sigma_{A}=-1,$ $\sigma_{B}=+1$. When the setting at $B$ is shifted to
$\theta_{B}^{\prime}=\pi/2$, the four quadrants again yield the same outcomes
as for $a_{A}=a_{B}$, there being still no crossing between quadrants. Thus
the right half still yields $\sigma_{A}=+1$ and the left half still yields
$\sigma_{A}=-1$, so there are no changes at $A$; but the top-right quadrant
now yields $\sigma_{B}=+1$ and the bottom-left $\sigma_{B}=-1$, so 50\% change
at $B$.]

This example proves that it is impossible to deduce a general lower bound on
$\alpha$ alone without assuming an exchange symmetry between the two wings. If
the measurement at $A$ is completed long before the measurement at $B$, the
degree of nonlocality from $B$ to $A$ tends to zero.

On the other hand, again in the limit $a_{A}/a_{B}\rightarrow\infty$, under a
shift $\theta_{A}=0\rightarrow\theta_{A}^{\prime}=-\pi/2$ at $A$ (keeping
$\theta_{B}=0$ fixed), it is found that a fraction $\beta(0,0,-\pi
/2)\rightarrow1/2$ change at $B$. Thus $\beta(0,0,-\pi/2)=\tilde{\beta
}(0,0,\pi/2)$, in accordance with rotational symmetry, and while the degree of
nonlocality is zero from $B$ to $A$ it is \textit{large} from $A$ to $B$,
saturating (3). [We may calculate the degree of nonlocality $\beta
(0,0,-\pi/2)$ from $A$ to $B$ for $a_{A}/a_{B}\rightarrow\infty$ by noting
that it must be the same as the degree of nonlocality $\alpha(0,0,\pi/2)$ from
$B$ to $A$ for $a_{B}/a_{A}\rightarrow\infty$. In the latter case, for
$\theta_{A}=0,\ \theta_{B}=0$ the branches $\psi_{+-}$ and $\psi_{-+}$ rapidly
separate along $r_{B}$, so that the top half of the initial square yields
$\sigma_{A}=-1,$ $\sigma_{B}=+1$ and the bottom half $\sigma_{A}=+1,$
$\sigma_{B}=-1$. With $\theta_{A}=0$, $\theta_{B}^{\prime}=\pi/2$, once again
the four quadrants yield the same outcomes as for $a_{A}=a_{B}$, and it is now
seen that 50\% have changed at $A$, so that for $a_{B}/a_{A}\rightarrow\infty$
we have $\alpha(0,0,\pi/2)\rightarrow1/2$. Thus we deduce that $\beta
(0,0,-\pi/2)\rightarrow1/2$ for $a_{A}/a_{B}\rightarrow\infty$.]

This last result is worth emphasising: if the coupling at $A$ is made much
larger than at $B$, the degree of nonlocality $\alpha$ from $B$ to $A$ becomes
small and the degree of nonlocality $\beta$ from $A$ to $B$ becomes large,
while the \textit{total} degree of nonlocality $\alpha+\beta=1/2$ is unchanged
(at least for angular shifts $\pm\pi/2$ and square pointer packets).

\section{Angular Settings Differing by $\pi$}

For measurements of spin, one must be careful to consider only those settings
$\theta_{A},\ \theta_{B}$ that correspond to physically distinct experimental
arrangements.\footnote{I am grateful to Lucien Hardy for raising this point.}
This can depend on how the measurements are carried out. For the interaction
Hamiltonian considered above, settings that differ by $\pi$ are physically
distinct; but in the case of Stern-Gerlach measurements, settings that differ
by $\pi$ (at one or both wings) must be identified with the same experiment.

To see this, it suffices to consider pilot-wave theory for a single spin. With
the Hamiltonian $\hat{H}=g(t)\hat{\sigma}_{\theta}(-i\partial/\partial r)$, if
the state $\left|  z+\right\rangle $ is measured for $\theta=0$ one finds that
all initial pointer positions $r(0)$ evolve towards $+\infty$; that is, one
always obtains $\sigma_{0}=+1$. If on the other hand one takes $\theta=\pi$,
all $r(0)$ evolve towards $-\infty$, corresponding to $\sigma_{\pi}=-1$. Thus,
for $\left|  z+\right\rangle $, a shift $\theta=0\rightarrow\theta=\pi$
induces a change in every outcome. The interaction is of course constructed so
that, assuming $g(t)\geq0$, the sign of the pointer position $r(t)$ at large
times tells us the value of the `observable' $\hat{\sigma}_{\theta}$. The
Hamiltonian changes sign $\hat{H}\rightarrow-\hat{H}$ under a shift
$\theta=0\rightarrow\theta=\pi$ (where $\hat{\sigma}_{\pi}=-\hat{\sigma}_{0}%
$), corresponding to a genuinely distinct experiment. For $\left|
z+\right\rangle $ this induces a reversal of the actual motion of the pointer.
Though of course, in every case the outcome corresponds to spin up along the
(fixed) \textit{z}-axis, as it must.

For $\left|  x+\right\rangle $, however, a shift $\theta=0\rightarrow
\theta=\pi$ leads to no change in outcome $\sigma_{\theta}$, corresponding to
a \textit{reversal} of spin along the \textit{z}-axis. To see this, simply
note that for both $\theta=0$ and $\theta=\pi$ half of the initial values
$r(0)$ must evolve towards $+\infty$ and half towards $-\infty$ (to yield the
correct outcome ratios); and since the trajectories $r(t)$ cannot cross (the
de Broglie-Bohm velocity field being single-valued), it follows that in both
cases $r(0)>0$ yields $r(t)\rightarrow+\infty$ ($\sigma_{0}=+1,\;\sigma_{\pi
}=+1$) and $r(0)<0$ yields $r(t)\rightarrow-\infty$ ($\sigma_{0}%
=-1,\;\sigma_{\pi}=-1$). Thus, while it may seem puzzling at first sight, in
this particular model an initial hidden-variable state that yields spin up
along \textit{z} for $\theta=0$ (that is, $\sigma_{0}=+1$) yields spin
\textit{down} along \textit{z} for $\theta=\pi$ (that is, $\sigma_{\pi}=+1$);
and similarly, spin down becomes spin up.\footnote{This seems strange from a
quantum viewpoint, because one tends to make the implicit and mistaken
assumption that so-called quantum `measurements' really do tell us the value
of something that existed beforehand. But contextuality tells us that, in
general, quantum `measurements' are not faithful -- that is, are not really
measurements, but simply experiments of a particular kind. Thus, there is no
reason why different `measurement' setups (with different Hamiltonians) could
not yield different values of spin along \textit{z} for the same
hidden-variable state, as in the example above.}

It is therefore clear that, for an interaction $\hat{H}=g(t)\hat{\sigma
}_{\theta}(-i\partial/\partial r)$, measurements along angles differing by
$\pi$ correspond to physically distinct experiments that can yield physically
distinct results. And so, it is not surprising that in our pilot-wave
calculations for the singlet state we found that a shift $\theta
_{B}=0\rightarrow\theta_{B}^{\prime}=\pi$ at $B$ can induce a large change in
the outcomes at $A$.

In contrast, for a Stern-Gerlach measurement on a single spin, the position of
the particle itself serves as a `pointer' [18]. And if the magnet is rotated
by $\pi$, this induces \textit{no} change in (the relevant part of) the
Hamiltonian\footnote{In an inhomogeneous magnetic field along the
\textit{z}-axis, we have a term $\hat{H}=\mu\hat{\sigma}_{z}B_{z}\simeq\mu
\hat{\sigma}_{z}(B_{z})_{z=0}+\mu\hat{\sigma}_{z}z(\partial B_{z}/\partial
z)_{z=0}$. Because of the second term, packets with opposite spins acquire
opposite \textit{z}-momenta and eventually separate [18]. If the magnet is
rotated by $\pi$, the sign of $(\partial B_{z}/\partial z)_{z=0}$ is
unchanged.}: the experimental arrangement is exactly the same, and therefore
the time evolution of the wavefunction and of the de Broglie-Bohm trajectories
must also be the same. Thus, for the state $\left|  z+\right\rangle $, the
motion of the `pointer' is \textit{not} affected by $\theta=0\rightarrow
\theta=\pi$. There is no physical distinction between the measurements with
$\theta=0$ and $\theta=\pi$, and in both cases the motion of the particle will
indicate spin up the \textit{z}-axis. Similarly, for $\left|  x+\right\rangle
$ no detail of the evolution is affected by $\theta=0\rightarrow\theta=\pi$:
one obtains spin up or down, depending on the initial particle position within
the initial packet, regardless of whether $\theta=0$ or $\theta=\pi$.

In the Stern-Gerlach case, then, settings that differ by $\pi$ correspond
physically to the same experiment. And so, for an entangled state of two spins
at $A$ and $B$, a shift $\theta_{B}=0\rightarrow\theta_{B}^{\prime}=\pi$ at
$B$ cannot induce any change at all at $A$. (There is, of course, no
contradiction with our pilot-wave calculations for the other case, which
corresponds to a different interaction.)

In the lower bounds (2)--(5), then, one must take into account the possibility
that some mathematically distinct settings may be physically identical. This
can be done by restricting the rotational angle $\delta$ to a range that does
not overcount the number of distinct experiments. Thus, in a Stern-Gerlach
case, where one must identify settings differing by $\pi$, (3) and (5) will
presumably be true -- if the appropriate symmetries hold -- for $\delta
\in(-\pi/2,+\pi/2)$, rotations larger than $\pm\pi/2$ being identified with
$\delta\mp\pi$. Then, in (5) for example, as the Stern-Gerlach magnet is
rotated from $\theta_{B}=0$ to $\theta_{B}^{\prime}=\pi/2$, the effect at $A$
will increase from $\alpha=0$ to $\alpha=0.25$; while further rotation --
actually corresponding to smaller values of $\left|  \delta\right|  $ -- will
decrease $\alpha$, and $\alpha=0$ when the magnet has rotated by $\pi$.

This point should be studied further; and a detailed comparison with
pilot-wave theory for the Stern-Gerlach case should be made.

\section{Subquantum Information}

We have defined the `total degree of nonlocality' $\alpha+\beta$ as the
equilibrium fraction $\alpha$ of outcomes at $A$ that change under a shift
$\theta_{B}\rightarrow\theta_{B}^{\prime}$ at $B$, plus the fraction $\beta$
of outcomes at $B$ that change under a shift $\theta_{A}\rightarrow\theta
_{A}^{\prime}$ at $A$ (whether the changes are from $+1$ to $-1$ or vice
versa). Clearly, $\alpha$ may be interpreted as the \textit{average number of
bits of information} per singlet pair transmitted nonlocally (in equilibrium)
from $B$ to $A$; and similarly for $\beta$ from $A$ to $B$. Thus $\alpha$ and
$\beta$ quantify the nonlocal flow of what might be termed `subquantum information'.

It has proved fruitful in recent years to look at quantum theory from an
information-theoretic perspective. While some of that work makes use of
insights from Bell's theorem, in the author's view the whole field of quantum
information would benefit from a more explicit hidden-variables (including
pilot-wave) perspective. Here are some examples.

\textit{Universal Lower Bound on Nonlocal Information Flow}: The lower bound
(2) on $\alpha(\theta_{A},\theta_{B},\theta_{B}+\delta)+\beta(\theta
_{A},\theta_{B},\theta_{A}-\delta)$ was derived assuming a rotational
symmetry. Would it be possible to derive a general lower bound on the total
nonlocal information flow in the singlet state, without this extra assumption?

The total degree of nonlocality $\alpha(\theta_{A},\theta_{B},\theta
_{B}^{\prime})+\beta(\theta_{A},\theta_{B},\theta_{A}^{\prime})$ depends on
the initial and final angular settings. It quantifies the total nonlocal flow
of information in the singlet state, generated by specific shifts $\theta
_{B}\rightarrow\theta_{B}^{\prime}$ and $\theta_{A}\rightarrow\theta
_{A}^{\prime}$ in the angular settings at $B$ and $A$. (That it depends on the
settings has been shown explicitly in the case of pilot-wave
theory.\footnote{We also saw that, individually, $\alpha$ and $\beta$ depend
on the measurement couplings, but their sum satisfies the lower bound (3)
obtained from rotational symmetry.}) Bell's theorem tells us that
$\alpha+\beta$ cannot vanish for all settings. But for a general
hidden-variables theory without any symmetries, there is no reason why
$\alpha+\beta$ could not be very small for some angles and very large for
others.\footnote{One can, for example, easily write down a local model that
reproduces the quantum correlations for the specific settings $\theta_{A}=0$,
$\theta_{B}=0$ and $\theta_{A}=0$, $\theta_{B}^{\prime}=\pi/2$ [1].} In other
words: the underlying nonlocality required by Bell's theorem could be
`concentrated' around certain ranges of settings. Thus it is not surprising
that we were able to derive a lower bound on $\alpha+\beta$ for
\textit{specific} settings only by assuming a rotational symmetry.

To circumvent this, one might consider the \textit{average} total degree of
nonlocality $\overline{\alpha+\beta}$ obtained by averaging $\alpha+\beta$
over all possible initial and final settings $\theta_{A},\theta_{B},\theta
_{A}^{\prime},\theta_{B}^{\prime}$. Alternatively, one might look at the
maximum value $(\alpha+\beta)_{\text{max}}$. It should be possible to derive a
general lower bound on $\overline{\alpha+\beta}$ or $(\alpha+\beta
)_{\text{max}}$ for the singlet state, without any extra symmetry assumption.
This would provide us with a universal lower bound on nonlocal information flow.

\textit{Degree of Nonlocality as a New Measure of Entanglement}: For a general
entangled pure state $\left|  \Psi\right\rangle $ of two spin-1/2 systems --
that is, for a pure bipartite state of two qubits -- various measures of
entanglement $E(\Psi)$ have been proposed in the literature. A recent example
is based on the decomposition $\left|  \Psi\right\rangle =p\left|  \Psi
_{e}\right\rangle +\sqrt{1-p^{2}}e^{i\phi}\left|  \Psi_{f}\right\rangle $ into
a maximally-entangled state $\left|  \Psi_{e}\right\rangle $ and an orthogonal
factorisable state $\left|  \Psi_{f}\right\rangle $, where $p$ and $\phi$ are
real; the measure $E(\Psi)\equiv p^{2}$ has been shown to be closely related
to a measure based on maximal violation of Bell's inequality [19]. Since
$\alpha+\beta$ -- the sum of the nonlocal effects from $B$ to $A$ and from $A$
to $B$ -- quantifies nonlocality, it is natural to ask if $\alpha+\beta$ can
also be used as a measure of entanglement.\footnote{I am grateful to Guido
Bacciagaluppi for this suggestion. (For a recent review of entanglement
measures, see for example the article by Leah Henderson in this volume.)}

One approach would be to use pilot-wave theory to calculate the angular
average $\overline{\alpha+\beta}$ or the maximum value $(\alpha+\beta
)_{\text{max}}$ for a general entangled state $\left|  \Psi\right\rangle $,
and to take $E(\Psi)=\overline{\alpha+\beta}$ or $E(\Psi)=(\alpha
+\beta)_{\text{max}}$. Such calculations could easily be performed
numerically, and the results compared with those derived from other
measures.\footnote{Should the result turn out to depend on the details of the
measurement process -- such as the shape of the pointer packets (assumed
square in the above) -- one could simply take the minimum value. (For $\alpha$
alone, of course, we have seen that even for the singlet state $\alpha$ may be
made arbitrarily small, in pilot-wave theory, by making the coupling at $A$
arbitrarily large; but the sum $\alpha+\beta$ is unaffected by this.)} Another
approach might be to try to derive theory-independent lower bounds on
$\overline{\alpha+\beta}$ or $(\alpha+\beta)_{\text{max}}$ for general
entangled states, and use the lower bounds as measures of entanglement.

Clearly, this proposal needs further study. Here, we restrict ourselves to
pointing out that, in the symmetric pilot-wave case with $a_{A}=a_{B}$ (and
with the same square pointer packets at $A$ and $B$), on the space of quantum
states $\alpha(0,0,\pi)$ is a local maximum for the maximally-entangled
singlet state. [It is straightforward to show that any infinitesimal
perturbation of the singlet state decreases $\alpha(0,0,\pi)$, recalling that
for the singlet we found $\alpha(0,0,\pi)=1/2$. First, for initial settings
$\theta_{A}=0,\ \theta_{B}=0$, if we write perturbed spin amplitudes%

\begin{align*}
a_{++}  &  =\epsilon_{++}\;,\;\;\;\ a_{+-}=\frac{1}{\sqrt{2}}+\epsilon_{+-},\\
\ a_{-+}  &  =-\frac{1}{\sqrt{2}}+\epsilon_{-+}\;,\;\;\;\ a_{--}=\epsilon_{--}%
\end{align*}
\bigskip with $\epsilon\equiv\sqrt{2}(\epsilon_{+-}+\epsilon_{+-}^{\ast
})=\sqrt{2}(\epsilon_{-+}+\epsilon_{-+}^{\ast})$ (the last equality following
from normalisation), and work to lowest order in $\epsilon$, the pointer
velocities differ from the corresponding singlet case only where $\psi_{+-}$
and $\psi_{-+}$ overlap: at such points $v_{A}=-v_{B}=a\epsilon$ (instead of
$v_{A}=v_{B}=0$). As a result, instead of obtaining $\sigma_{A}=-1,$
$\sigma_{B}=+1$ and $\sigma_{A}=+1,$ $\sigma_{B}=-1$ for points respectively
above and below the line $r_{B}=r_{A}$, those values are obtained for points
respectively above and below the line $r_{B}=(1+\epsilon)/(1-\epsilon
)r_{A}+\epsilon\Delta/(1-\epsilon)$ for $-\Delta/2\leq r_{A}\leq
-\epsilon\Delta/2$ and $r_{B}=(1-\epsilon)/(1+\epsilon)r_{A}+\epsilon
\Delta/(1+\epsilon)$ for $-\epsilon\Delta/2\leq r_{A}\leq+\Delta/2$.
Similarly, with the new settings $\theta_{A}=0,\ \theta_{B}^{\prime}=\pi$ the
same (perturbed) state has the spin amplitudes%

\begin{align*}
a_{++}^{\prime}  &  =\frac{1}{\sqrt{2}}+\epsilon_{+-}\;,\;\;\;\ a_{+-}%
^{\prime}=-\epsilon_{++},\\
\ a_{-+}^{\prime}  &  =\epsilon_{--}\;,\;\;\;\ a_{--}^{\prime}=\frac{1}%
{\sqrt{2}}-\epsilon_{-+}%
\end{align*}
and $v_{A}^{\prime}=v_{B}^{\prime}=a\epsilon$ in the overlap of $\psi
_{++}^{\prime}$ and $\psi_{--}^{\prime}$, yielding $\sigma_{A}=+1,$
$\sigma_{B}=+1$ and $\sigma_{A}=-1,$ $\sigma_{B}=-1$ for points respectively
above and below the line $r_{B}=-(1+\epsilon)/(1-\epsilon)r_{A}-\epsilon
\Delta/(1-\epsilon)$ for $-\Delta/2\leq r_{A}\leq-\epsilon\Delta/2$ and
$r_{B}=-(1-\epsilon)/(1+\epsilon)r_{A}-\epsilon\Delta/(1+\epsilon)$ for
$-\epsilon\Delta/2\leq r_{A}\leq+\Delta/2$. Simple geometry then shows that
the fractional area of points in the initial square for which outcomes have
changed at $A$, under $\theta_{B}=0\rightarrow\theta_{B}^{\prime}=\pi$, is now
equal to%
\[
\alpha(0,0,\pi)=\frac{1}{2}-\frac{5}{4}\epsilon^{2}+O(\epsilon^{3})
\]
which is a local maximum at $\epsilon=0$. It remains to be seen if
$\alpha(0,0,\delta)$ has a local maximum at $\epsilon=0$ for all angles
$\delta$.]

\textit{Classical Simulation of Entanglement}: We have said that $\alpha$
($\beta$) is equal to the average number of bits of subquantum information per
singlet pair transmitted nonlocally, in equilibrium, from $B$ to $A$ ($A$ to
$B$). In other words, $\alpha$ ($\beta$) is the average amount of information
per pair that needs to be transmitted faster than light from $B$ to $A$ ($A$
to $B$) in order to reproduce the EPR-correlations. It might be interesting to
investigate how this is related to recent work on the simulation of quantum
entanglement with classical communication [20].\footnote{Averaging (5) over
$\delta\in(-\pi,\;\pi)$, we obtain a mean lower bound of 0.25 bits; if instead
only $\delta\in(-\pi/2,\;\pi/2)$ count as physically distinct settings, the
mean lower bound is $\frac{1}{4}(1-\frac{2}{\pi})\simeq0.09$ bits. One expects
the results would be higher without the symmetry assumptions on which (5) is based.}

\textit{Subquantum Computation}: In pilot-wave theory it is straightforward to
show that the hidden-variable trajectory of a particle can contain information
corresponding to \textit{all} the results of a parallel quantum computation --
for example if the particle is guided by a superposition of overlapping energy
eigenfunctions whose eigenvalues encode the results of the computations (where
here the computations are `performed' by the evolution of the pilot wave in
configuration space, and the results are `read' by the piloted particle). This
information could be read by us if we had access to matter in a state of
quantum nonequilibrium $\rho\neq\left|  \psi\right|  ^{2}$, leading to a truly
exponential speed-up in processing power not just for some problems, but quite
generally [5, 10, 21].

\section{Conclusion and Hypothesis}

Summarising, we have demonstrated a general `signal-locality theorem', which
states that in any deterministic hidden-variables theory that reproduces
quantum statistics for some `equilibrium' distribution $\rho_{eq}(\lambda)$ of
hidden variables $\lambda$, a generic `nonequilibrium' distribution
$\rho(\lambda)\neq\rho_{eq}(\lambda)$ would give rise to instantaneous signals
at the statistical level (as occurs in pilot-wave theory). Further, for an
equilibrium ensemble of EPR-experiments, assuming certain symmetries we have
derived lower bounds on the fraction of systems whose outcomes change under a
shift in the distant measurement setting, and we have verified that the bounds
are satisfied by pilot-wave theory. We have also pointed out some potential
benefits of a perspective based on `subquantum information'.

With the signal-locality theorem in hand, let us now consider what its
physical implications might be.

Bell's theorem is widely regarded as proving that if hidden variables exist
then so do instantaneous influences. But there is no consensus on what to
conclude from this. A widespread conclusion is that hidden variables do not
exist, the `argument' being that relativity would otherwise have to be
violated. But this is rather like someone in 1901 arguing that atoms cannot
exist because if they did Newtonian mechanics would have to be violated. There
is no reason why known physical principles cannot be violated at some
hitherto-unknown level.\footnote{Though there is disagreement about the status
of relativity even among those who do consider that hidden variables might
exist: some try to construct nonlocal theories in which the symmetries of
Minkowski spacetime are somehow preserved at the fundamental level, while
others (including this author) propose that special relativity be abandoned,
with Minkowski spacetime emerging only as an equilibrium phenomenology. See
ref. [10].}

It is important, then, to distinguish between Bell's theorem and what various
authors have concluded from it. Similarly, one must distinguish between the
signal-locality theorem proved above and what this author proposes to conclude
from it.

The author suggests that the signal-locality theorem has the following
physical significance: it indicates that our universe is in a special
`finely-tuned' state in which statistical noise happens to precisely mask the
effects of nonlocality -- a state of statistical equilibrium which is not
fundamental but merely contingent.

For it seems mysterious that nonlocality should be hidden by an all-pervading
quantum noise, and that (as we have shown) any deviation from that noise would
make nonlocality visible. It is as if there is some sort of `conspiracy' in
the laws of physics that prevents us from using nonlocality for signalling.
But the apparent conspiracy evaporates if one recognises that our universe is
in a state of statistical equilibrium at the hidden-variable level, a special
state in which nonlocality \textit{happens} to be hidden.

On this view, the physics we see is not fundamental; it is merely a
phenomenological description of an equilibrium state [15]. Fundamentally, the
universe is nonlocal, obeying laws that have yet to be uncovered (pilot-wave
theory providing a possible example). Unfortunately our experience is confined
to an equilibrium state that hides the true nature of things behind a veil of
quantum noise. Since any small departure from equilibrium would reveal the
underlying nonlocal physics, our present inability to observe nonlocality
directly (as opposed to indirectly, via Bell's theorem) is not enforced by any
fundamental physical principle: it is merely a contingent feature of equilibrium.

Indeed, our general inability to control the hidden-variable level is a
contingent feature of equilibrium. This is clear in pilot-wave theory, where
the uncertainty principle holds if and only if $\rho=|\psi|^{2}$ [15]. And it
may be shown that the same is true in any deterministic hidden-variables
theory: the uncertainty principle is valid if and only if $\rho(\lambda
)=\rho_{eq}(\lambda)$ [10]. From this perspective, immense practical resources
-- for communication, and also for computation -- are hidden from us by a veil
of uncertainty noise, because we happen to live in quantum equilibrium.

It is then natural to make the hypothesis that the universe began in a state
of quantum nonequilibrium $\rho(\lambda)\neq\rho_{eq}(\lambda)$, where
nonlocal signalling was possible and the uncertainty principle was violated,
the relaxation $\rho(\lambda)\rightarrow\rho_{eq}(\lambda)$ taking place
during the great violence of the big bang [5, 10, 14--16]. As this relaxation
occurred, the possibility of nonlocal signalling faded away, and statistical
uncertainty took over. The quantum equilibrium state $\rho(\lambda)=\rho
_{eq}(\lambda)$ may then be seen, heuristically, as a kind of `quantum heat
death' -- analogous to the classical thermodynamic heat death in which all
systems in the universe have reached the same temperature. (In the classical
heat death, thermal energy may no longer be used to do work; in the `quantum
heat death', the underlying nonlocality may no longer be used for signalling.)
In effect we are suggesting that, some time in the remote past, a
hidden-variables analogue of the classical heat death actually occurred in our universe.

One might also compare our present limitations with those of a Maxwell demon
in thermal equilibrium with a gas, whose attempts to sort fast and slow
molecules fail. On this view the common objection to hidden variables -- that
their detailed behaviour can never be observed -- is seen to be misguided: for
the theory can hardly be blamed if we happen to live in a state of statistical
equilibrium that masks the underlying details. There is no reason why
nonequilibrium $\rho(\lambda)\neq\rho_{eq}(\lambda)$ could not exist in the
remote past or in distant regions of the universe [5, 10, 14, 15], in which
case the details of the `hidden-variable level' would not be hidden at all.

It may seem gratuitous to draw such radical conclusions from what has been
proven here. But our hypothesis may also be supported by arguments from other
areas of physics.

The view that our universe is in an equilibrium state is arguably supported by
quantum field theory in curved spacetime, where there is no clear distinction
between quantum and thermal fluctuations [22]. On this basis it has in fact
been argued by Smolin that quantum and thermal fluctuations are really the
same thing [23]. This suggests that quantum theory is indeed just the theory
of an equilibrium state, analogous to thermal equilibrium.

This view is strengthened by Jacobson's demonstration that the Einstein field
equations may be viewed as an `equation of state', in the context of a
`thermodynamics of spacetime' [24]. According to Jacobson, quantum field
fluctuations correspond to some sort of `equilibrium' distribution that might
be violated at high energies; here, we would interpret quantum fluctuations as
corresponding to $\rho(\lambda)=\rho_{eq}(\lambda)$, which may have been
violated at early times.

Further support comes from cosmology. The notorious `horizon problem' has been
with us for decades: the cosmic microwave background is observed to be nearly
isotropic, and yet, at the time when photons decoupled from matter, the
observable universe supposedly consisted of a large number of causally
disconnected regions.\footnote{Though it should be pointed out that the
calculation of the size of causal horizons assumes that the classical
Friedmann expansion, with scale factor $\propto t^{1/2}$ at early times, is
valid all the way back to $t=0$. It might equally be argued that the horizon
problem is an artifact of this (rather naive) assumption.} Inflation was
thought to avoid this, but more recent analysis shows that at least some
inflationary models merely shift the problem to having to assume `acausal'
homogeneity as an initial condition in order to obtain inflation [25]. A
number of workers have seen in the horizon problem a hint that some sort of
superluminal causation is required at early times, whether by topological
fluctuations that lead to an effective nonlocality [26], or by the more recent
suggestion that the speed of light increases at high energies [27]. Our
hypothesis offers another alternative: if the universe started in quantum
nonequilibrium $\rho(\lambda)\neq\rho_{eq}(\lambda)$, the resulting nonlocal
effects may have played a role in homogenising the universe at early times [10].

But is there any prospect of testing these ideas experimentally?
Investigations are proceeding on two fronts. First, if we accept the idea from
inflation that the temperature fluctuations in the microwave background were
ultimately seeded by quantum fluctuations at very early times, then precise
measurements of the microwave background can be seen as probes of quantum
theory in the very early universe: if $\rho(\lambda)\neq\rho_{eq}(\lambda)$
during the inflationary era, this would leave an imprint on the microwave
background that differs from the one predicted by standard quantum field
theory [10, 17]. Second, certain exotic (perhaps supersymmetric) particles may
have decoupled at very early times, before quantum equilibrium was reached; if
so, they may still exist in a state of quantum disequilibrium today and they
-- or their decay products -- would now violate quantum mechanics [10, 14, 16, 17].

\textbf{Acknowledgements.} For helpful comments and discussions I am grateful
to Guido Bacciagaluppi, Jossi Berkovitz, Lucien Hardy, Lee Smolin and
Sebastiano Sonego, and to audiences at the Universities of Maryland, Notre
Dame and Utrecht. I would also like to thank Jeremy Butterfield and Tomasz
Placek for the invitation to present this work in the magnificent city of
Cracow, to express my gratitude to Tadeusz (`Tadzio') Litak for kindly
initiating me into Cracow's more intimate and magical mysteries, and to offer
my best wishes for the development of foundations of physics in Central and
Eastern Europe. This work was supported by the Jesse Phillips Foundation.

\bigskip\bigskip

\begin{center}
\textbf{REFERENCES}
\end{center}

\bigskip

[1] Bell, J.S. (1964) On the Einstein-Podolsky-Rosen paradox, \textit{Physics}
\textbf{1}, 195--200.

[2] de Broglie, L. (1928) La Nouvelle Dynamique des Quanta, in J. Bordet
\textit{et al.} (eds.), \textit{\'{E}lectrons et Photons: Rapports et
Discussions du Cinqui\`{e}me Conseil de Physique}, Gauthier-Villars, Paris,
105--141. [English translation: Bacciagaluppi, G. and Valentini, A.
(forthcoming) \textit{Electrons and Photons: The Proceedings of the Fifth
Solvay Congress}, Cambridge University Press, Cambridge.]

[3] Bohm, D. (1952) A suggested interpretation of the quantum theory in terms
of `hidden' variables. I and II, \textit{Physical Review} \textbf{85},
166--179; 180--193.

[4] Bell, J.S. (1987) \textit{Speakable and Unspeakable in Quantum Mechanics},
Cambridge University Press, Cambridge.

[5] Valentini, A. (1992) On the pilot-wave theory of classical, quantum, and
subquantum physics, PhD thesis, International School for Advanced Studies,
Trieste, Italy.

[6] Holland, P. (1993) \textit{The Quantum Theory of Motion: an Account of the
de Broglie-Bohm Causal Interpretation of Quantum Mechanics}, Cambridge
University Press, Cambridge.

[7] Bohm, D. and Hiley, B. J. (1993) \textit{The Undivided Universe: an
Ontological Interpretation of Quantum Theory}, Routledge, London.

[8] Cushing, J. T. (1994) \textit{Quantum Mechanics: Historical Contingency
and the Copenhagen Hegemony}, University of Chicago Press, Chicago.

[9] Cushing, J. T. \textit{et al.} (eds.) (1996) \textit{Bohmian Mechanics and
Quantum Theory: an Appraisal}, Kluwer, Dordrecht.

[10] Valentini, A. (forthcoming)\textit{ Pilot-Wave Theory of Physics and
Cosmology}, Cambridge University Press, Cambridge.

[11] Bell, J. S. (1966) On the problem of hidden variables in quantum
mechanics, \textit{Reviews of Modern Physics} \textbf{38}, 447--452.

[12] Kochen, S. and Specker, E. P. (1967) The problem of hidden variables in
quantum mechanics, \textit{Journal of Mathematics and Mechanics} \textbf{17}, 59--87.

[13] Valentini, A. (1991) Signal-locality, uncertainty, and the subquantum
\textit{H}-theorem. I, \textit{Physics Letters A} \textbf{156}, 5--11.

[14] Valentini, A. (2001) Hidden Variables, Statistical Mechanics and the
Early Universe, in J. Bricmont \textit{et al}. (eds.), \textit{Chance in
Physics: Foundations and Perspectives}, Springer, Berlin.

[15] Valentini, A. (1991) Signal-locality, uncertainty, and the subquantum
\textit{H}-theorem. II, \textit{Physics Letters A} \textbf{158}, 1--8.

[16] Valentini, A. (1996) Pilot-Wave Theory of Fields, Gravitation and
Cosmology, in ref. [9].

[17] Valentini, A. (forthcoming) Hidden variables, quantum fluctuations, and
the early universe, \textit{International Journal of Modern Physics A}.

[18] Bohm, D. (1951) \textit{Quantum Theory}, Prentice-Hall, New York.

[19] Abouraddy, A. F. \textit{et al}. (2001) Degree of entanglement for two
qubits, \textit{Physical Review A} \textbf{64}, 050101.

[20] Brassard, G. \textit{et al}. (1999) Cost of exactly simulating quantum
entanglement with classical communication, \textit{Physical Review Letters}
\textbf{83}, 1874--1877; Cerf, N.J. \textit{et al}. (2000) Classical
teleportation of a quantum bit, \textit{Physical Review Letters} \textbf{84},
2521--2524; Steiner, M. (2000) Towards quantifying non-local information
transfer: finite-bit non-locality, \textit{Physics Letters A} \textbf{270},
239--244; Massar, S. \textit{et al}. (2001) Classical simulation of quantum
entanglement without local hidden variables, \textit{Physical Review A}
\textbf{63}, 052305.

[21] Valentini, A. (forthcoming) Subquantum Information and Computation, in
\textit{Proceedings of the Second Winter Institute on Foundations of Quantum
Theory and Quantum Optics}.

[22] Sciama, D. W., Candelas, P. and Deutsch, D. (1981) Quantum field theory,
horizons and thermodynamics, \textit{Advances in Physics} \textbf{30}, 327--366.

[23] Smolin, L. (1986) On the nature of quantum fluctuations and their
relation to gravitation and the principle of inertia, \textit{Classical and
Quantum Gravity} \textbf{3}, 347--359.

[24] Jacobson, T. (1995) Thermodynamics of spacetime: the Einstein equation of
state, \textit{Physical Review Letters} \textbf{75}, 1260--1263.

[25] Vachaspati, T. and Trodden, M. (2000) Causality and cosmic inflation,
\textit{Physical Review D} \textbf{61}, 023502.

[26] Hawking, S. W. (1982) The Boundary Conditions of the Universe, in H.A.
Br\"{u}ck \textit{et al}. (eds.), \textit{Astrophysical Cosmology}, Vatican
City, 563--574.

[27] Moffat, J. W. (1993) Superluminary universe: a possible solution to the
initial value problem in cosmology, \textit{International Journal of Modern
Physics D} \textbf{2}, 351--366; Clayton, M. A. and Moffat, J. W. (1999)
Dynamical mechanism for varying light velocity as a solution to cosmological
problems, \textit{Physics Letters B}\textbf{ 460}, 263--270; Albrecht, A. and
Magueijo, J. (1999) Time varying speed of light as a solution to cosmological
puzzles, \textit{Physical Review D} \textbf{59}, 043516; Barrow, J. D. (1999)
Cosmologies with varying light speed, \textit{Physical Review D} \textbf{59}, 043515.
\end{document}